\documentclass[conference]{IEEEtran}

\usepackage{graphicx}
\usepackage{url}
\usepackage{amsmath}
\usepackage{amsthm, amssymb}

\usepackage{tikz}
\usetikzlibrary{patterns}
\usetikzlibrary{shapes,arrows}
\usetikzlibrary{circuits.ee.IEC}
\usepackage[utf8]{inputenc}
\usepackage{cite}

\newtheorem{lem}{Lemma}
\newtheorem{rem}{Remark}

\IEEEoverridecommandlockouts

\begin{document}

\title{Performance of Media-based Modulation in Multi-user Networks \\ \huge{(Invited Paper) }\thanks{The work of Merve Y\"{u}zge\c{c}cio\u{g}lu has received partly funding from the European Union's Horizon 2020 research and innovation programme under the Marie Sklodowzka-Curie grant agreement No 641985}}

\author{\IEEEauthorblockN{Merve Y\"{u}zge\c{c}cio\u{g}lu}
\IEEEauthorblockA{Communications Laboratory\\
TU Dresden\\
D-01062 Dresden, Germany\\
Email: merve.yuzgeccioglu@tu-dresden.de}
\and
\IEEEauthorblockN{Eduard Jorswieck}
\IEEEauthorblockA{Communications Laboratory\\
TU Dresden\\
D-01062 Dresden, Germany\\
Email: eduard.jorswieck@tu-dresden.de}}

\maketitle

\begin{abstract}
High spectral efficiency and low power consumption are the most challenging requirements of $5$G networks since the number of devices are increased drastically. Media-based modulation (MBM) is a promising scheme in order to achieve these requirements. In this paper, the spectral efficiency of MBM scheme in single- and multi-user networks is studied. To find the spectral efficiency, information theoretical methods are used and tight lower- and upper-bounds are derived for the achievable rate. The performance of the system is studied for both correlated and uncorrelated constellation diagrams. It is shown that the entries of the constellation diagram for the sum rate region of multi-user case are always correlated. The characteristic of covariance matrix is derived and its impact on performance assessed. 
\end{abstract}

\IEEEpeerreviewmaketitle

\section{Introduction}
In future wireless communication scenarios a huge number of devices will communicate tremendous amounts of data with low latencies and high energy efficiencies \cite{Andrews2012}. In order to achieve these goals, novel multi-antenna techniques as massive multiple-input multiple-output (MIMO), higher carrier frequencies as mmWave and novel modulation schemes as generalized frequency division multiplexing (GFDM) are considered. Often, one disadvantage of the technological approaches is that with increased hardware (e.g. number of radio frequency (RF) chains in massive MIMO) and complexity, the energy consumption is significantly increased. Therefore, we consider a recently proposed modulation scheme in \cite{Khandani2013} and study its applicability for multi-user scenarios. 

The most common way in wireless communications to send information is to choose symbols from a complex modulation alphabet and transmit data over a fading channel. In other words, data is modulated before leaving the transmitter. In contrast to traditional methods, complex channel coefficients may be used to convey information. A method to modulate data after leaving the transmitter is  introduced in \cite{Khandani2013} called MBM. The concept of MBM is generating a constellation diagram by perturbing the multipath fading channel. In this case, the constellation size is related to the number of perturbations as $2^{N_P} = M$ where  $N_P$ is the number of the channel perturbations and $M$ is the resulting constellation size. For each transmission, one of the channel realizations is chosen to send information and the receiver is able to decode received data due to the training period at the beginning of each coherence time of the channel. 

Recently, the spectral efficiency of a Rayleigh fading single-input multiple-output (SIMO) channel is investigated and it is shown in \cite{Khandani2013} and \cite{Khandani2014} that by using MBM, the capacity of additive white Gaussian noise (AWGN) channel is achieved asymptotically. Since the coefficients of the fading channel are used to transmit data, training of the receiver is critical. This leads to the high complexity of the pilot training and afterward the decoding period in high data rates. To overcome these issues, so called layered multiple-input multiple-output MBM (LMIMO-MBM) is introduced  and a recursive decoding algorithm for receiver side is presented \cite{Khandani2015}. Furthermore,  the performance of LMIMO-MBM is investigated in \cite{Xu2016} with secrecy constraint. In \cite{Naresh2016}, bit error rate performance of MBM is examined when implemented to different physical layer techniques.  A new technique to design reconfigurable antenna systems which promises low power and low cost that can be used to perturb the channel is introduced in \cite{Boutayeb2014}. Another approach to modulate information for MBM by using time-varying plasma is presented and experimental results are shown in \cite{Yang2015}. Since the entries of the constellation diagram are generated by using channel state information (CSI), increasing the constellation size in MBM increase the complexity much more than the conventional modulation methods. In order to overcome this drawback, two channel state selection algorithms are proposed in \cite{Yapeng2016}. In \cite{Chen2016}, polar coded MBM is introduced to achieve a better performance. Another scheme called differential MBM is given in \cite{Naresh2017} where a low-complexity  maximum-likelihood detector is also proposed. 

In this paper, we consider the single-user link with correlated constellation diagram entries and observe that correlation decreases the achievable spectral efficiency. We have observed that for the multiple-access channel (MAC), even if the entries of  the constellation diagrams of each user are uncorrelated, the resulting constellation diagram of sum rate region is correlated. The covariance matrix for arbitrary number of users $K$ and modulation points $M$ is computed. Interestingly, the null space grows with increasing $K$ and $M$. The achievable rate region is compared to AWGN region and with the ergodic capacity region. It is shown that for asymptotic $M$, the impact of correlation vanishes. 

The structure of the paper is as follows. In Section \ref{sec:SystemModel}, the system models of the single- and multi-user network are introduced. Achievable rate performance of the system is studied in Section \ref{sec:PerformanceAnalysis} and the numerical results are shown in \ref{sec:NumericalResults}. Finally, paper is concluded in \ref{sec:Conclusion}.

\section{System Model}\label{sec:SystemModel}
\subsection{Single user link}
In MBM scheme, there are $M$ constellation points which are constructed from the channel coefficients by using different kind of hardware elements (RF mirrors, parasitic antennas, etc.). Transmitter chooses one of the constellation points to transmit the intended message. The resulting system model can be written as 

\begin{equation}
y = \sqrt{P}\mathbf{x}(m) + n,
\end{equation}

where $\mathbf{x}\in \mathbb{C}^{M\times 1}$ contains the symbols from the constellation diagram with $M$ correlated entries, $\mathbf{x}(m)$ is the $m$-th entry of the constellation diagram. Since we consider a single-input single-output (SISO) system, $\mathbf{x}(m)$, $y$ and $n$ are scalars and $P$ is the transmit power. $\mathbf{x}$ has Gaussian distribution with $\mathcal{N}$($0, \mathbf{\Sigma}$) and the AWGN noise has $\mathcal{N}$($0, \sigma^2$). As a result, the probability density function of the output signal $y$ will be in form of a Gaussian mixture

\begin{equation}
f_{Y}(y) =  \frac{1}{M}\sum\limits_{m=1}^M\frac{1}{\sqrt{2\pi}\sigma}\exp\left\{-\frac{[y-\sqrt{P}\mathbf{x}(m)]^2}{2\sigma^2}\right\}.
\end{equation}

\subsection{Multiple access channel}
In this section, a $K$-user MAC in Rayleigh fading environment is considered. Each user chooses one of the channel realizations to transmit the intended information and maps each message $m_k$ to the corresponding codeword $\mathbf{x}_k(m_k)$ where $m = 1,...,M$ and $k = 1,...,K$. The input signals $\mathbf{x}_k(m_k)$ are chosen from the constellation diagram with $M$ entries where $\mathbf{x}_k$ has Gaussian distribution with $\mathcal{N}$($0, \mathbf{\Sigma}$). The input-output relationship of a MAC is 
\begin{equation}
y = \sqrt{P}\sum\limits_{k=1}^K\mathbf{x}_k(m_k) + n,
\end{equation}
where $m = 1,...,M$ and $n$ is the AWGN with $\mathcal{N}$($0, \sigma^2$).  Due to $M$ point constellation diagram with Gaussian components, the resulting distributions of the output $y$ is 
\begin{equation}\label{eq:ypdf}
\begin{split}
f_{Y}(y) &=  \frac{1}{M^K}\sum\limits_{m_1=1}^M\dots\sum\limits_{m_K=1}^M\frac{1}{\sqrt{2\pi}\sigma}\\
&\quad\cdot \exp\left\{-\frac{[y-\sqrt{P}(\mathbf{x}_1(m_1)+\dots+\mathbf{x}_K(m_K))]^2}{2\sigma^2}\right\}
\end{split}
\end{equation}
and the conditional pdf $y|\mathbf{x}_k$ is
\begin{equation}\label{eq:ycondpdf}
f_{Y|\mathbf{X}_k}(y|\mathbf{x}_k) =  \frac{1}{M}\sum\limits_{m=1}^M\frac{1}{\sqrt{2\pi}\sigma}\exp\left\{-\frac{[y-\sqrt{P}\mathbf{x}_k(m_k)]^2}{2\sigma^2}\right\}.
\end{equation}

We observe from Eq.~[\ref{eq:ypdf}] that there are $M^K$ constellation points $\mathbf{x}_1(m_1)+\dots+\mathbf{x}_K(m_K)=\mathbf{x}(m_1,\dots,m_K)$ for $1\leq m_k\leq M$ and $1\leq k \leq K$. These points are found by summation of each possibility among $K$ users with $M$ messages.

\section{Performance Analysis}\label{sec:PerformanceAnalysis}
In this section, achievable rate region of the singe-user link is studied for correlated constellation diagram entries. For the MAC case, both correlated and uncorrelated entries are taken into account in order to study the behavior of the system.

\subsection{Single-user link}

The achievable rate of the system can be calculated by using the information theoretical methods

\begin{equation}
I(\mathbf{x}(m);y) = h(y) - h(n).
\end{equation}

Here, the noise component has a Gaussian distribution with $\sigma^2$ variance, it is straight forward that the entropy can be found as $h(n) = \frac{1}{2}\log_2(2\pi e\sigma^2)$. Since the received signal has Gaussian mixture distribution, calculation of the entropy of $y$ is not as easy as the noise component, as a matter of fact there exists no closed form solution for the entropy of a Gaussian mixture. A tight approximation is introduced to literature recently in \cite{Moshksar2016}. By using this approximation method, entropy of the received signal can be bounded as
\begin{equation}
(\gamma+\alpha_{N,N'}) \log_2e \leq h(y) \leq (\gamma+\beta_{N,N'}) \log_2e. \label{eq:entropy}
\end{equation}
The detailed explanation of the parameters can be found in \cite{Moshksar2016}. 

\subsubsection{Impact of the correlation}
Depending on the hardware realization of the channel perturbations, correlation between the entries of $\mathbf{x}(m)$ can occur. It is well known from the analysis of the multi-antenna links that the spatial correlation lowers the achievable rate, if the transmitter has no CSI and the receiver has perfect CSI, by applying majorization theory \cite{Jorswieck2007}. 

The entropy of a correlated Gaussian vector $\mathbf{x}$ is a Schur-concave function of the eigenvalues of the correlation matrix $\mathbf{\Sigma}$, i.e., if
\begin{equation}\label{eq:Majorization}
\lambda (\mathbf{\Sigma}_1) \succeq \lambda (\mathbf{\Sigma}_2) \quad \mathrm{then} \quad 
h(\mathbf{\Sigma}_1) \leq h(\mathbf{\Sigma}_2), 
\end{equation}
with the majorization inequality $\mathbf{x} \succeq \mathbf{y}$ meaning $\sum\limits_{m=1}^l \mathbf{x}_{[m]}\geq \sum\limits_{m=1}^l \mathbf{y}_{[m]}$, $1 \leq l \leq M-1$, and $\sum\limits_{m=1}^M \mathbf{x}_m=\sum\limits_{m=1}^M \mathbf{y}_m$, and entropy $h(\mathbf{\Sigma}_1) = \log (2\pi e)^M \det(\mathbf{\Sigma}_1)$.

The identity correlation matrix $\mathbf{\Sigma}=\mathbf{I}$ is majorized by all other correlation matrices and provides the highest entropy. Since this result holds for all symmetric and concave functions of the correlation matrix $\mathbf{\Sigma}$, we expect a similar behavior for the  MBM link and discuss the impact of correlation based on the numerical simulations in the next section.

\subsection{Multiple access channel}
The achievable rate region of a $K$ user multiple access channel is given by
\begin{equation} \label{eq:MAC}
\begin{split}
\{R_\mathcal{S} &< I(\mathbf{x}_\mathcal{S}(m_\mathcal{S});y|\mathbf{x}_{\mathcal{S}^c}(m_{\mathcal{S}^c}))\\
& \quad \forall \mathcal{S} \subseteq \mathcal{K} = \{1,2,\dots,K\}\},
\end{split}
\end{equation}
with $R_\mathcal{S} = \sum\limits_{k \in \mathcal{S}} R_k$ and $\mathbf{x}_\mathcal{S}(m_\mathcal{S}) = \{\mathbf{x}_k(m_k)\}_{k\in \mathcal{S}}$ and $\mathcal{S}^c =\mathcal{K}\setminus \mathcal{S}$. The mutual information to define sum rate  can be written as
\begin{equation}
I(\mathbf{x}_1(m_1),\dots,\mathbf{x}_K(m_K);y) = h(y) - h(n).
\end{equation}

Again the entropy of the noise $n$ can be written as $h(n) = \frac{1}{2}\log_2(2\pi e\sigma^2)$ and for the remaining term  $h(y)$ upper and lower bounds are found as in Eq.~[\ref{eq:entropy}]. Similarly, the other mutual information expressions in (\ref{eq:MAC}), i.e., $I(\mathbf{x}_\mathcal{S}(m_\mathcal{S});y|\mathbf{x}_{\mathcal{S}^c}(m_{\mathcal{S}^c}))$ can be derived. 

\subsubsection{Covariance matrix analysis}

Interestingly, even if the original constellation diagram has uncorrelated entries the resulting signal set with $M^K$ entries will be correlated. The covariance matrix $\mathbf{\Sigma}$ can be constructed by using the below description 

\begin{equation}\label{eq:covarianceentries}
    \mathbb{E}\{\mathbf{x}_i(m_n)\mathbf{x}_j(m_l)\}= 
\begin{cases}
    \rho_{n,l},& \text{if } i = j \quad \text{and} \quad n\not= l\\
    1,& \text{if } i = j \quad \text{and} \quad n = l\\
		0,& \text{if } i \not= j, 
\end{cases}
\end{equation}

where $\rho_{n,l}$ is the correlation coefficient of the entries of the $k-th$ user where $k = 1,...,K$  and it is $0$ for uncorrelated case.  It can be seen from Eq.~[\ref{eq:covarianceentries}] that the joint constellation diagram of the users with uncorrelated constellation points will have a covariance matrix $\mathbf{\Sigma}$ which is different than the identity matrix $\mathbf{I}_{M^K \times M^K}$. Furthermore, the correlation matrix majorizes the identity matrix $$\mathbf{\Sigma} \succeq \mathbf{I}_{M^K \times M^K} \succeq \mathbf{0}.$$

Note that, a similar analysis can be performed for other multi-user communication scenarios, too. For both, the broadcast and the interference channel setup, similar superposition of the individual constellation points result in a correlation in the joint constellation diagram of the sum rate signal set. 

The basic assumption in \cite{Khandani2013} is that entries of the constellation diagram are i.i.d. Gaussian zero-mean random variables. However, this is not true for the joint constellation diagram of a $K$-user MAC. Furthermore, the covariance matrix of the system can be defined as $\mathbf{\Sigma}=\mathbb{E}\{\mathbf{x}(m_1,\dots,m_K)\mathbf{x}(m_1,\dots,m_K)^T\}$. To see this effect clearly, covariance matrix of a $2$-user MAC with $M$ points constellation diagram is given 

\begin{equation}
\mathbf{\Sigma}=\begin{bmatrix}
    \mathbf{D}_{M\times M}  & \mathbf{I}_{M\times M} & \dots  & \mathbf{I}_{M\times M} \\
    \vdots & \vdots  & \ddots & \vdots \\
    \mathbf{I}_{M\times M} & \mathbf{I}_{M\times M} & \dots  & \mathbf{D}_{M\times M}
\end{bmatrix}
\end{equation}

\[
\mathrm{with}\quad \mathbf{D}=\begin{bmatrix}
    2 &1  & \dots  & 1 \\
    \vdots  & \vdots & \ddots & \vdots \\
    1 & 1  & \dots  & 2
\end{bmatrix}.
\]

The cases with more than $2$ users and the correlated channel coefficients can be derived similarly by using Eq.~[\ref{eq:covarianceentries}]. However, it is tedious and results in a complicated structured covariance matrix. 	Nevertheless, it is possible to characterize the eigenvalues of these covariance matrices.

\begin{lem}
The covariance matrix $\mathbf{\Sigma}$ has $M^K$ eigenvalues. The largest eigenvalue is $KM^{K-1}$, there are $K(M-1)$ eigenvalues which are $M^{K-1}$ and the remaining eigenvalues are zero. 
\end{lem}

\begin{rem}
There is a significant number of zero eigenvalues in the covariance matrix $\mathbf{\Sigma}$, e.g. for $K = 3$, $M = 3$, there are $20$ zeros out of $27$ eigenvalues. 
\end{rem}

\begin{figure}[!t]
\centering
\includegraphics[scale=0.4]{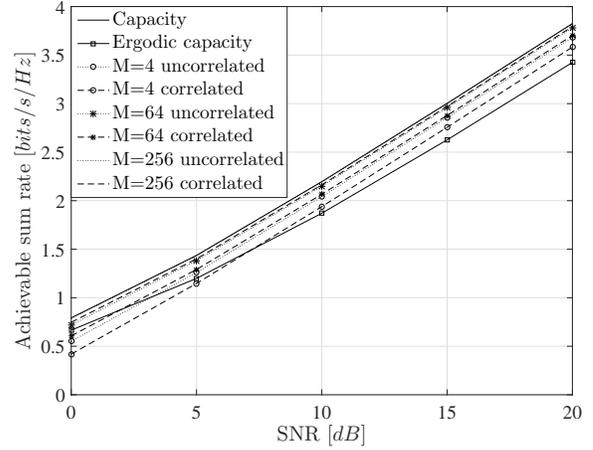}
\caption{Sum rate behavior of 2-user MAC with various size constellation diagrams.}
\label{MACcorrelation}
\end{figure}

\begin{rem}
 However, when $M$ goes infinity, the non-zero eigenvalues will be dominant and the effect of the correlation will vanish. This can be seen in Fig.~\ref{MACcorrelation}. Here, the sum rate of a $2$-user MAC is considered. In three different cases each user have $2$, $4$ and $16$ points uncorrelated constellation diagrams, respectively. For the resulting constellation diagrams with $4$, $16$ and $256$ entries, degradation causing from the correlation gets smaller with the increased size of constellation diagram.
\end{rem}

\section{Numerical results}\label{sec:NumericalResults}
\begin{figure}[!t]
\centering
\includegraphics[scale=0.4]{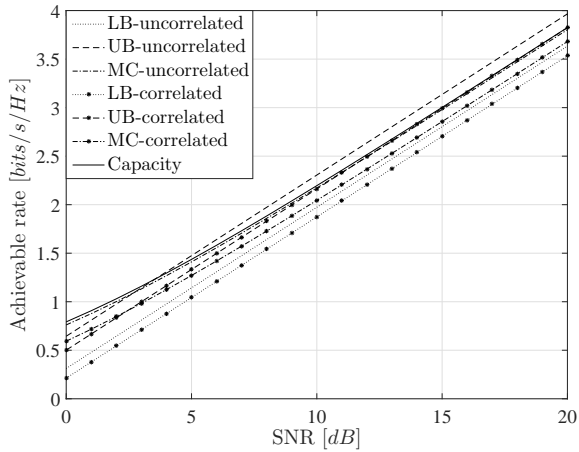}
\caption{Achievable rate of MBM scheme with uncorrelated and heavily correlated constellation diagram and $N_P = 8$ channel perturbations}
\label{single8}
\end{figure}

In this section, numerical results for both single- and multi-user system is presented. To have an insight into the behavior of the singe-user system with MBM scheme under correlation, achievable rate results for the $N_P=8$, $M=256$ points constellation diagram with correlated and uncorrelated channel coefficients are shown. In Fig.~\ref{single8}, the numerical results for upper (UB) and lower bounds (LB) of the achievable rate are provided and compared with the Monte Carlo (MC) simulation. The capacity curve corresponds with the AWGN capacity. Even though the number of constellation points is considerably low, the channel capacity is almost achieved with MBM scheme. In the case with strong correlation among constellation points, a degradation on achievable rate was expected as stated in Eq.~[\ref{eq:Majorization}].  However, the degradation causing from the correlation is considerably low. Thus, we can observe that the behavior of the MBM scheme with high number of $M$ under correlated channel is similar to uncorrelated case.  It can be seen from Fig.~\ref{single8} that the degradation causing from a strong correlation between constellation points with $M=256$ is only $0.1$ bits/s/Hz. Observe that all curves have the same slope, e.g., the same degrees of freedom. 

\begin{figure}[!t]
\centering
\includegraphics[scale=0.4]{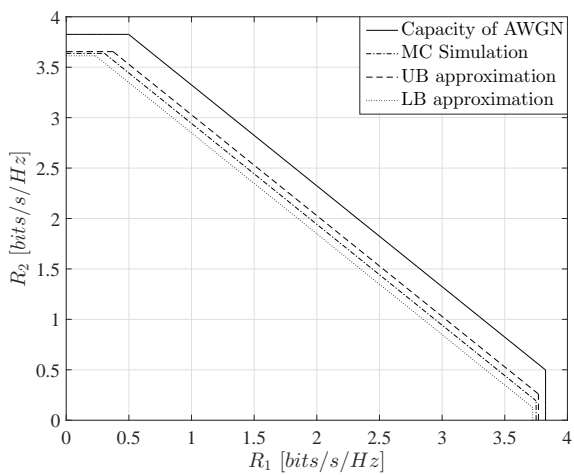}
\caption{Achievable rate region for a $2$-user MAC system with  $N_P=3$ channel perturbations at $20dB$ SNR}
\label{Nrf3SNR20}
\end{figure}

Furthermore, the performance of a $2$-user MAC with $3$ channel perturbations is shown in Fig.~\ref{Nrf3SNR20}. The achievable rate region of the system is close to the capacity of the $2$-user MAC system in AWGN channel with Gaussian codebook. It is seen from this figure that the gap between AWGN capacity and the MBM scheme is a little bit wider for the sum rate than the individual rates of the users. This is the effect of inherent correlation of joint constellation diagram as stated in Eq.~[\ref{eq:covarianceentries}]. Overall, the gap between the achievable rate region of the MBM scheme and the AWGN capacity region is small even with low number of channel perturbations. Numerical results show that with implementation of MBM scheme in a MAC system, communication in high data rates is possible without consuming additional power.

\section{Conclusion}\label{sec:Conclusion}
In this work, the achievable rate performance of MBM in both single- and multi-user network is studied. The effect of the correlated constellation diagram in a single user link is investigated. We have shown that the effect of the correlation on the spectral efficiency is considerably low. When the constellation diagram is expanded, degradation caused by the correlation vanishes, eventually capacity of the channel is achieved. Furthermore, the achievable rate region of the MAC is derived. Tight approximation of the achievable rate for both cases is derived and the tightness of the approximation is shown.

In future, this work may be extended to analyze the asymptotic behavior of MBM achievable rate for correlated constellation points when $M \to \infty$. Spectral efficiency of MIMO-MBM as described in \cite{Khandani2015} may be studied to investigate the impact of the spatial correlation on the achievable rate. And finally, model verification based on realistic RF perturbation approaches, e.g. RF mirrors \cite{Khandani2013,Khandani2014,Khandani2015} can be studied.

%\printbibliography
\bibliographystyle{IEEEtran}
\bibliography{ref}
%\flushend
\end{document}